\documentclass[lettersize,journal]{IEEEtran}
\usepackage{amsmath,amsfonts}
\usepackage{algorithm}
\usepackage{array}
\usepackage[caption=false,font=normalsize,labelfont=sf,textfont=sf]{subfig}
\usepackage{textcomp}
\usepackage{algorithm}
\usepackage{algpseudocode}
\usepackage{enumitem}
\usepackage[english]{babel}
\usepackage{amsthm}
\usepackage{multicol}
\usepackage{color}
\usepackage{physics}
\usepackage{graphics}

\newtheoremstyle{mystyle}
  {\topsep}
  {\topsep}
  {} 
  {}          
  {\bfseries} 
  {.}         
  { }         
  {}         

\theoremstyle{mystyle}
\newtheorem{theorem}{Theorem}
\newtheorem{Lemma}{Lemma}

\theoremstyle{remark}
\newtheorem*{remark}{Proof}
\usepackage{physics}
\usepackage{stfloats}
\usepackage{amssymb}
\usepackage{commath}
\usepackage{upgreek}
\usepackage{amsmath}
\DeclareMathOperator{\diag}{diag}

\usepackage{url}
\usepackage{verbatim}
\usepackage{multicol}
\usepackage{lipsum}
\usepackage{graphicx}
\usepackage{cite}
\begin{document}


\title{Optimal Operation of Active RIS-Aided Wireless Powered Communications in IoT Networks}

\author{Waqas Khalid, Alexandros-Apostolos A. Boulogeorgos, \textit{Senior Member, IEEE}, Trinh Van Chien, \textit{Member, IEEE}, Junse Lee, \textit{Member, IEEE}, Howon Lee, \textit{Senior Member, IEEE}, Heejung Yu, \textit{Senior Member, IEEE}

\thanks{This research was supported by ``Regional Innovation Strategy (RIS)" through the National Research Foundation of Korea (NRF) funded by the Ministry of Education (MOE) (2021RIS-004), by the Basic Science Research Program through the NRF funded by the MOE (NRF-2022R1I1A1A01071807, 2021R1I1A3041887), by the Institute of Information \& communications Technology Planning \& Evaluation (IITP) grant funded by the government of Korea (Ministry of Science \& ICT (MSIT)) (2022-0-00704, Development of 3D-NET Core Technology for High-Mobility Vehicular Service). A.-A. A. Boulogeorgos research was supported by MINOAS project, within the H.F.R.I call ''Basic Research Financing (Horizontal Support of all Sciences)'' under the National Recovery and Resilience Plan “Greece 2.0” funded by the European Union - NextGenerationEU (H.F.R.I. Project Number: 15857). \textit{(Corresponding authors: Howon Lee and Heejung Yu.)}}

\thanks{Waqas Khalid is with the Institute of Industrial Technology, Korea University, Sejong 30019, Korea. Email: waqas283@\{korea.ac.kr, gmail.com\}.}
\thanks{A.-A. A. Boulogeorgos is with the Department of Electrical and Computer Engineering, University of Western Macedonia, 50100 Kozani, Greece. Email: al.boulogeorgos@ieee.org.}
\thanks{Trinh Van Chien is with the School of Information and Communication Technology, Hanoi University of Science and Technology, Hanoi 100000, Vietnam. Email: chientv@soict.hust.edu.vn.}
\thanks{Junse Lee is with the School of AI Convergence, Sungshin Women's University, Seoul 02844, Korea. Email: junselee@sungshin.ac.kr}
\thanks{Howon Lee is with the Department of Electrical and Computer Engineering, Ajou University, Suwon, Gyeonggi 16499, Korea. Email: howon@ajou.ac.kr.}
\thanks{Heejung Yu is with the Department of Electronics and Information Engineering, Korea University, Sejong 30019, Korea. Email: heejungyu@korea.ac.kr.}}



\maketitle
\begin{abstract} 
Wireless-powered communications (WPCs) are increasingly crucial for extending the lifespan of low-power Internet of Things (IoT) devices. Furthermore, reconfigurable intelligent surfaces (RISs) can create favorable electromagnetic environments by providing alternative signal paths to counteract blockages. The strategic integration of WPC and RIS technologies can significantly enhance energy transfer and data transmission efficiency. However, passive RISs suffer from double-fading attenuation over RIS-aided cascaded links. In this article, we propose the application of an active RIS within WPC-enabled IoT networks. The enhanced flexibility of the active RIS in terms of energy transfer and information transmission is investigated using adjustable parameters. We derive novel closed-form expressions for the ergodic rate and outage probability by incorporating key parameters, including signal amplification, active noise, power consumption, and phase quantization errors. Additionally, we explore the optimization of WPC scenarios, focusing on the time-switching factor and power consumption of the active RIS. The results validate our analysis, demonstrating that an active RIS significantly enhances WPC performance compared to a passive RIS.
\end{abstract}


\section{Introduction}
Future wireless communication systems will support Internet of Things (IoT) networks, enabling interactions among diverse devices via wireless channels and Internet gateways, with a focus on self-sustainability to minimize maintenance costs \cite{R0, R0a, R0b, R0c}. Self-powered wireless devices can harvest energy from external sources, such as sun, wind, and vibrations. However, their performance is often dependent on natural factors, such as weather conditions. Recently, radio frequency (RF)-based power transfer has gained considerable attention due to its adaptability, flexibility, wide coverage, and low-complexity implementation \cite{R1}. Wireless-powered communication (WPC) represents a green networking paradigm that integrates wireless information transmission and power transfer \cite{R2}. In WPC systems, devices achieve self-sustainability by efficiently harvesting energy from RF signals, thereby reducing the need for frequent battery replacement. This approach enhances throughput, extends battery life, and lowers operational costs \cite{R3}. The primary objective of WPC systems is to enhance energy transfer efficiency over long distances while ensuring reliable data transmission \cite{R1,R2,R3,R3a}.

Recently, reconfigurable intelligent surfaces (RISs) have emerged as an innovative physical layer technology, specifically designed to refine wireless channels and create an intelligent radio environment \cite{R6}. RISs employ reflective elements to dynamically adjust the phase and amplitude of incoming signals, thereby addressing propagation challenges, such as blockage, fading, and shadowing, without relying on complex encoding/decoding or intensive signal processing \cite{R7}. By managing electromagnetic (EM) wave scattering and reflection, RISs significantly reduce the impact of radio frequency (RF) impairments and signal distortion, eliminating the need for additional base stations or relay nodes, thereby substantially enhancing coverage and wireless connectivity. Fundamentally, an RIS transforms the radio environment from a passive to an active element in optimization strategies \cite{R8}. This innovation allows for the optimization of transmitters, receivers, and channels, leading to significant improvements in data rates, reduced latency, enhanced reliability, and increased energy and spectral efficiency. By employing strategic placement and advanced beamforming techniques, an RIS can enhance signal quality, reduce power consumption, mitigate interference, and ensure reliable signal reception \cite{R9,R10}.

\subsection{Motivation}
WPC is expected to play a crucial role in supporting low-power and energy-limited IoT devices in the upcoming sixth-generation (6G) era. However, optimizing power transfer and data transmission across dynamic environments, characterized by challenging radio propagation over long distances, necessitates innovative solutions \cite{R11}. In the literature, advanced techniques for enhancing network performance typically involve extensive RF chains for data transmission over high-frequency bands, which increase energy consumption and hardware costs \cite{R6,R7,R8,R9,R10,R11}. In this context, RISs are critical for improving coverage, capacity, and transmission reliability, thereby significantly boosting the spectral and energy efficiency of WPC in IoT networks \cite{R11a}. The inherent adaptability of RISs enhances network connectivity, ensuring stable communications by boosting data rates and mitigating distance-related channel attenuation. While a passive RIS offers minimal noise, leading to high array gain, it also faces challenges, such as double-fading attenuation over RIS-aided cascaded links. To address this issue, conventional solutions involve deploying numerous passive elements, which increases surface size and circuit power consumption \cite{R13}, making them unsuitable for IoT devices. Typically, the path loss over the transmitter--RIS--receiver link exceeds that of a direct link, limiting performance improvements, such as capacity gain, especially in scenarios with a strong direct link. Consequently, a passive RIS, limited to phase adjustment without amplification, suffers from low received signal power due to compound path loss \cite{R12,R12a,R12b}. 

Active RIS technology can effectively address these limitations \cite{R14}. Active RISs mitigate double-fading attenuation by employing low-power reflection-type amplifiers in each element\footnote{Active RISs employ power amplifiers and phase-shift circuits, thereby eliminating the need for digital-to-analog and analog-to-digital converters as well as mixers, which are required in active relays. This design reduces hardware costs and power consumption \cite{R15}.}. Amplifying signals at individual elements reduces the number of active elements required to achieve the desired signal-to-noise ratio (SNR) at the receiver, marking a significant advancement in RIS technology \cite{R14,R15}. However, further investigation into power amplification by an active RIS is required to enhance power transfer and data transmission in energy-constrained IoT devices, focusing on the adjustable parameters of an active RIS-aided WPC.

\subsection{Contribution and Organization}
In this article, we investigate the integration of an active RIS to enhance WPC within IoT networks. An active RIS employs power amplification to boost the power transfer and data transmission capabilities of IoT devices with limited energy resources. Our study delves into the potential of an active RIS in WPC, emphasizing its tunable parameters. Specifically, we address the non-negligible active noise and power consumption intrinsic to the amplification process in an active RIS. Our comprehensive analysis includes an in-depth examination of channel behavior, providing a thorough evaluation of the ergodic rate and outage probability. This evaluation meticulously considers critical factors, such as signal amplification, active noise, power consumption, and phase quantization errors. Furthermore, we optimize WPC scenarios by addressing the time-switching factor and power consumption of an active RIS. The results validate the effectiveness of our analytical framework and demonstrate significant performance enhancements compared to systems employing a passive RIS. 

The remainder of this article is organized as follows. Section~\ref{Sec2} introduces the system model, including network description, channel modeling, active RIS model, phase quantization error, and the procedures for energy transfer and information transmission. Section~\ref{Sec3} presents the performance analysis, including ergodic rates and outage probability, along with the power consumption of an active RIS. Section~\ref{Sec4} details the optimization of the ergodic and effective rates, focusing on the time-switching factor and power consumption. Section~\ref{Sec5} provides numerical results that demonstrate the effectiveness of an active RIS in WPC. Finally, Section~\ref{Sec6} concludes the article and highlights future work.

$Notations$: Vectors and matrices are written in boldface, with matrices in capitals. All vectors are column vectors. For a matrix $\mathbf{A}$, $\mathbf{A}^H$ denotes the conjugate transpose of $\mathbf{A}$. $\diag(x_1, \cdots, x_n)$ denotes a diagonal matrix with diagonal elements $x_1, \cdots, x_n$. For a vector $\mathbf{x}$, $|\mathbf{x}|$ and $||\mathbf{x}||$ refer to the element-wise absolute value and Euclidean norm of $\mathbf{x}$, respectively. $\mathcal{CN}(\mu, \sigma^2)$ represents a complex Gaussian distribution with mean $\mu$ and variance $\sigma^2$. $\mathcal{U}[ \alpha, \beta ]$ denotes the uniform distribution between $\alpha$ and $\beta$. $\mathcal{GM}\left(s,r\right)$ represents a Gamma distribution with shape $s$ and scale $r$. Additionally, $\mathbb{E}\left\{.\right\}$ and $\mathbb{V}\left\{.\right\}$ denote the statistical expectation and variance, respectively. The Gamma function is defined as $\Gamma\left(z\right) = \int_{0}^{\infty} t^{z-1}e^{-t}dt$.

\begin{figure*}[t!]
\centering
\includegraphics[width=7in,height=3.6in]{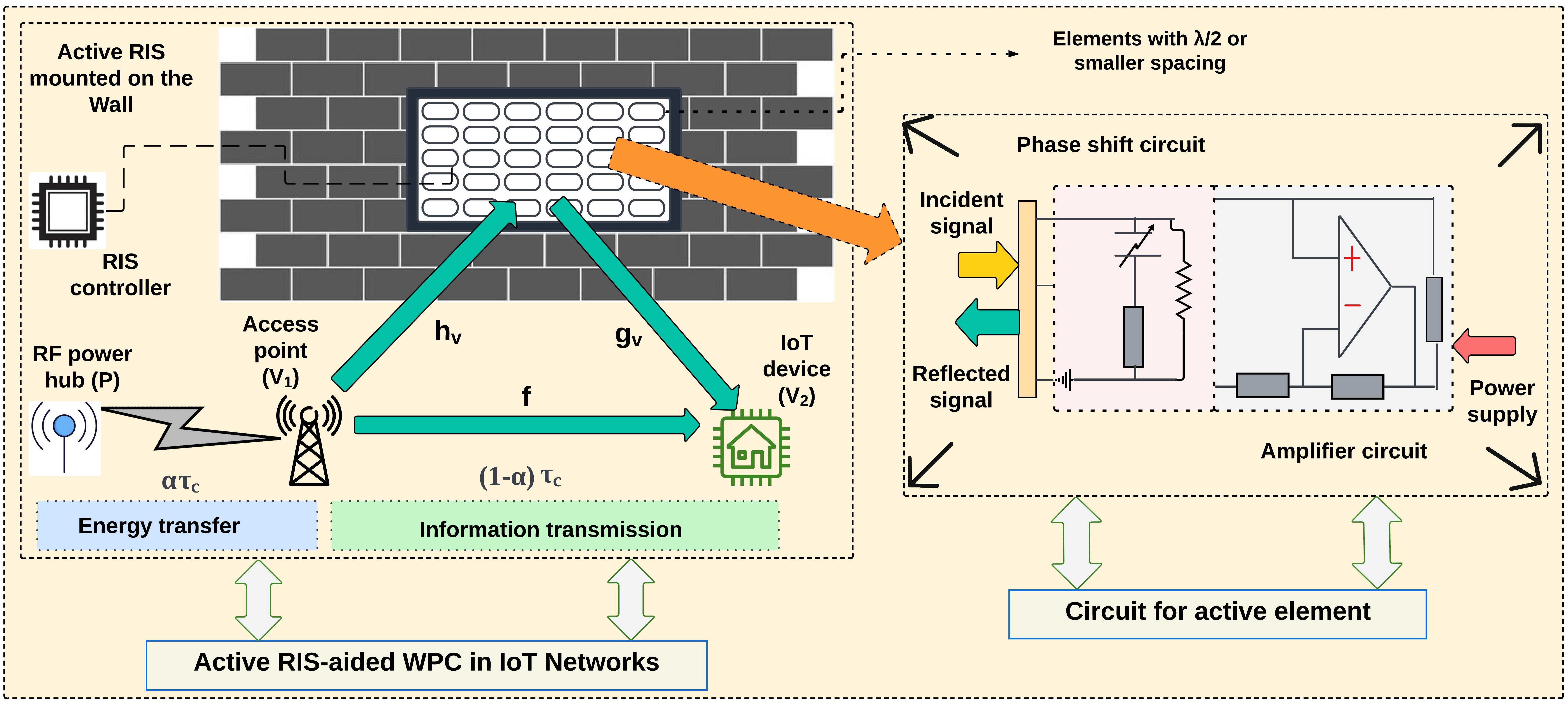}
\caption {System model diagram of active RIS-aided WPC in IoT networks, including the circuit configuration of active elements.}
\label{diagram1}
\end{figure*}

\section{System Model} \label{Sec2}

\subsection{Network Description} We consider a typical WPC system for IoT devices, as illustrated in Fig. 1.  The system comprises a wireless RF power hub ($P$) and two IoT devices: $V_1$, which functions as an access point, and $V_2$\footnote{Each node in our system is equipped with a single antenna. Future research will explore the potential of multi-antenna configurations and beamforming techniques to further improve system performance.}. A strategically positioned RIS with $M$ elements is located between $V_1$ and $V_2$ and employs active beamforming to fine-tune the amplitude and phase, thereby enhancing information transmission. This RIS-WPC framework serves as a foundational model for understanding the integration of an active RIS with WPC, highlighting the synergy between energy transfer and information transmission \cite{R17}. The energy-constrained $V_1$ effectively harvests RF energy from $P$ and uses this power to transmit information to $V_2$. We assume that the RIS elements are spaced at half the wavelength $\frac{\lambda}{2}$ (where $\lambda$ is the wavelength) to ensure independent fading channels at $V_2$. Both direct and RIS-aided indirect links are utilized between $V_1$ and $V_2$ to meet stringent signal-to-interference-plus-noise ratio (SINR) requirements at $V_2$ \cite{R18}. 

\subsection{Channel Modeling} The small-scale fading coefficients between $P$ and $V_1$, and between $V_1$ and $V_2$, are denoted as $h_p=\sqrt{\zeta_p}\hat{h}_p$ and $f=\sqrt{\zeta_f}\hat{f}$, respectively. Additionally, the small-scale fading vectors from $V_1$ to RIS and from RIS to $V_2$ are represented by $\mathbf{h}_v=[h_{1},...,h_{m},...,h_{M}]^H$ and $\mathbf{g}^H_v=[g_{1},...,g_{m},...,g_{M}]$, respectively, where $h_{m}=\sqrt{\zeta_{h_m}}\hat{h}_{m}$ and $g_m=\sqrt{\zeta_{g_m}}\hat{g}_m$.  The path-loss coefficients, $\zeta_p=d^{-\epsilon}_{p}$, $\zeta_f=d^{-\epsilon}_{f}$, $\zeta_{h_m}=d^{-\epsilon}_{h_m}$, and $\zeta_{g_m}=d^{-\epsilon}_{g_m}$  are determined by the path loss exponent $\epsilon$ and the distances $d_{p}$, $d_{f}$, $d_{h_m}$, and $d_{g_m}$ from $P$ to $V_1$, from $V_1$ to $V_2$, from $V_1$ to the $m$th RIS element, and from the $m$th RIS element to $V_2$, respectively. All channels undergo independent block Rayleigh fading\footnote{This modeling is representative of urban environments where structures often scatter radio signals, making the line-of-sight links uncertain and justifying the assumption of Rayleigh fading for indirect RIS links \cite{R19}.}. In this article, we assume perfect channel state information (CSI), enabling us to establish an upper bound on performance for our proposed framework\footnote{For channel estimation, several efficient schemes designed for RIS-aided communications can be adapted \cite{R1a}. While perfect CSI serves as a theoretical benchmark, addressing the practical scenario of imperfect CSI is essential. In this context, channel estimation error can be modeled as a Gaussian random variable with a defined variance that quantifies the impact of imperfect CSI and serves as an additional noise factor \cite{R1b}. However, considering the specific scope and contributions of this study, this aspect is reserved for future research.}. The small-scale fading coefficients follow Gaussian distributions \cite{R20}: $\hat{f}\sim \mathcal{CN}\left(0,1\right)$, 
$\hat{h}_p \sim \mathcal{CN}\left(0,1\right)$, 
$\hat{h}_m \sim \mathcal{CN}\left(0,1\right)$ and
$\hat{g}_m \sim \mathcal{CN}\left(0,1\right)$.

\subsection{Active RIS Model} In our design, we incorporate power amplifiers and phase-shifting circuits for individual RIS elements. This configuration allows precise signal control, encompassing reflection and amplification, to mitigate the considerable path loss over RIS-aided links and effectively combat the inherent double-fading challenge \cite{R21}. The reflecting coefficient matrix (reflect beamforming) of the active RIS is given by $\mathbf{\Theta}=\diag\{\phi_1,...,\phi_M\}$, where $\phi_m = \rho_m e^{j\theta_m}$ and $\forall m = 1,..., M$ represents the reflecting coefficient of the $m^{th}$ element and $\theta_m$ and $\rho_m$ denote the phase shift and amplification factor, respectively. Each amplifier is assigned a unique amplification factor, $0\leq\rho_m\leq\rho_m^{*}$, offering flexible beamforming design \cite{R22}, where $\rho_m^{*}$ is the predetermined maximum amplitude.

In this article, we assume that the processing delays introduced by the active RIS are negligible. This assumption is based on the fact that processing delays in active RISs are on the order of nanoseconds, while the propagation delays of EM waves are at the microsecond level. Moreover, advancements in active RIS technology have incorporated low-latency signal processing techniques explicitly designed to minimize such delays. Therefore, these delays remain minimal and have an insignificant impact on system performance \cite{R1c}.

\subsection{Phase Error} 
Achieving significant beamforming gain with an RIS requires leveraging accurate CSI across all channels, which substantially enhances the reception quality at $V_2$. This enhancement is made possible by advanced RIS-aided channel estimation methods, which facilitate precise phase shift adjustments ($\theta^{*}_m$) to optimally align the RIS phase with the effective channel, considering both direct and RIS-cascaded links \cite{R23}. However, hardware limitations restrict the phases $\left\{\theta_m\right\}^{M}_{m=1}$ of the RIS elements to a finite set of discrete values, known as quantized phase shifting \cite{R24}. Consequently, the RIS controller selects the nearest quantized phase for the $m^{th}$ element from the set $\mathcal{F}=\left\{0,\frac{2\pi}{2^b},...,\frac{\left(2^b-1\right)2\pi}{2^b}\right\}$, where $b$ represents the number of quantization bits. The selection function $\mathrm{f_1}\left(\theta^{*}_m\right)$ maps $\theta^{*}_m$ to the closest phase in $\mathcal{F}$, satisfying $\mathrm{f_1}\left(\theta^{*}_m\right)=\theta^{e}_m$ under the condition that $|\theta^{*}_m-\theta^{e}_i|\leq |\theta^{*}_m-\theta^{e}_j|$, where $\theta^{e}_i$, $\theta^{e}_j \in \mathcal{F}, \forall i \neq j$. Furthermore, $\phi^e_m=\mathrm{f_1}\left(\theta^{*}_m\right)-\theta^{*}_m$ represents the phase quantization error of the $m^{th}$ element, which tends toward a uniform distribution as the number of quantization levels increases,  with $\phi^e_m=~\mathcal{U}[-\pi 2^{-b},\pi 2^{-b})$ \cite{R25}.

\subsection{Energy Transfer and Information Transmission} In each time interval, denoted by $\tau_c$, the communication process unfolds in two sequential stages: energy transfer and information transmission. Initially, for a duration of $\alpha \tau_c$, where $\alpha$ denotes the time-switching factor, $V_1$ harnesses the energy from the wireless RF power hub $P$. This energy harvesting phase results in an accumulated energy $E_H$, given by $E_H=\eta \alpha \tau_c P_{p} |h_p|^2$, where $\eta$ is the efficiency of the rectenna for RF-to-direct current (DC) conversion \cite{R26} and $P_{p}$ is the transmission power of $P$. Subsequently, during the remaining time $(1-\alpha)\tau_c$, $V_1$  uses the harvested energy $E_H$ to transmit information to $V_2$. The transmission power employed by $V_1$ for information transmission is calculated as follows:
\begin{align}\label{equ:1}
P_{v}=\frac{E_{H}}{\left(1-\alpha\right)\tau_c}=\frac{\eta \alpha P_{p} |h_p|^2}{\left(1-\alpha\right)}
\end{align}

The signal reaches $V_2$ through both direct and RIS-aided links\footnote{Active RISs with nanosecond processing delays handle the incident and reflected signals carrying the same symbol within a single timeslot. Due to non-ideal inter-element isolation, the active RIS might receive part of the reflected signals which results in feedback-type self-interference \cite{R1d}. To account for self-interference, we consider a self-interference channel matrix $\mathbf{Q}$. The received signal, including self-interference, can be modeled by modifying equation (2). This involves replacing the RIS reflection matrix, $\mathbf{\Theta}$, with the equivalent matrix $\mathbf{\Theta}_{eq} = \left(\mathbf{I_M-\mathbf{\Psi}  \mathbf{Q}}\right)^{-1}\mathbf{\Psi}$, where $\mathbf{\Psi}$ denotes the RIS reflecting matrix in the non-ideal case, incorporating self-interference. Therefore, the analytical approaches in this article can be extended to active-RIS systems with self-interference by adopting the equivalent RIS reflecting matrix ($\mathbf{\Theta}_{eq}$) instead of the actual one ($\mathbf{\Theta}$). This study does not consider self-interference, as its primary goal is to establish a baseline understanding of system performance without the added complexity introduced by self-interference, leaving this aspect for future work.}, represented as follows:

\begin{align}
y_v&=\underbrace{\sqrt{P_{v}}f x_v}_{\text{direct link}}+\underbrace{\mathbf{g}^H_v \mathbf{\Theta}\left(\sqrt{P_{v}} \mathbf{h}_vx_v+\mathbf{z}_v \right)}_{\text{RIS-aided link}}+n_v\nonumber \\
&=\underbrace{\sqrt{P_{v}}\left(f+\mathbf{g}^H_v \mathbf{\Theta}\mathbf{h}_v\right)x_v}_{\text{Desired signal}}+\underbrace{\mathbf{g}^H_v \mathbf{\Theta}\mathbf{z}_v}_{\text{Noise for active RIS}} +\underbrace{n_v} _{\text{Noise for $V_2$}}
\label{equ:2}
\end{align} 
where $x_v$ represents the transmitted signal for $V_2$ and is an independent complex Gaussian random variable with a zero mean and unit variance. Furthermore, $\mathbf{z}_v \sim \mathcal{CN}( \mathbf{0}, \sigma_v^2 \mathbf{I})$ is the noise introduced by the active RIS, which is non-negligible because of amplification by amplifiers \cite{R27}. Additionally, $n_v \sim \mathcal{CN}\left(0,\sigma_{n}^2\right)$ represents the static noise at $V_2$ unrelated to $\mathbf{\Theta}$.

Considering that the received signal in Eq. \eqref{equ:2}, the SINR for $V_2$ can be written as:
\begin{align}
\gamma_{v}=\;\;&\frac{P_{v}|f+\mathbf{g}^H_v \mathbf{\Theta}\mathbf{h}_v|^2}{\sigma_{v}^2\|\mathbf{g}^H_v \mathbf{\Theta}\|^2+\sigma_{n}^2} \nonumber \\
=\;\;&\frac{\nu_1 |h_p|^2 |f+\mathbf{g}^H_v \mathbf{\Theta}\mathbf{h}_v|^2}{\sigma_{v}^2\|\mathbf{g}^H_v \mathbf{\Theta}\|^2+\sigma_{n}^2}
\label{equ:3}
\end{align}
where $\nu_1=\frac{\eta \alpha P_{p}}{\left(1-\alpha\right)}$.

By incorporating the phase quantization error for the active RIS, the SINR received at $V_2$ can be written as:
\begin{align}
\gamma_{v}\; \;=\;\;\frac{\nu_1 |h_p|^2 \left| |f|+  \sum_{m=1}^{M}\rho_m|g_m||h_m| e^{j\phi^e_m}\right|^2}{\sigma_{v}^2\|\mathbf{g}^H_v\mathbf{\Theta}\|^2+\sigma_{n}^2}
\label{equ:4}
\end{align}

\section{Performance Analysis}\label{Sec3}

In this section, we introduce a theoretical framework for evaluating the ergodic rate and outage probability as performance measures. In addition, we evaluate the power consumption of the active RIS.

\subsection{Ergodic Rate}

With Eq. \eqref{equ:4}, an ergodic rate can be defined as follows:

\begin{align}
R_v=\left(1-\alpha\right) \mathbb{E} \left\{ \log_2\left(1+\gamma_{v}\right) \right\}.
\label{equ:5}
\end{align}
Because only $(1-\alpha)$ portion of the total communication interval is used for information transmission, the achievable rate is scaled by $(1-\alpha)$ in Eq. \eqref{equ:5}. To derive the tractable expression of Eq. \eqref{equ:5}, its upper bound is obtained as follows:
\begin{align}
\hat{R}_v=\left(1-\alpha\right)\log_2 \left(1+ \frac{\nu_1 \mathbb{E}\left\{T_1 +T_2\right\}}{\mathbb{E}\left\{T_3+\sigma_{n}^2\right\}} \right)
\label{equ:6}
\end{align}
where $T_1=|h_p|^2\left| |f|+  \sum_{m=1}^{M}\rho_m |g_m||h_m| \cos{\left(\phi^e_m\right)}\right|^2 $,\; $T_2=|h_p|^2\left|  \sum_{m=1}^{M}\rho_m  |g_m||h_m| \sin{\left(\phi^e_m\right)}\right|^2$ and $T_3=\sigma_{v}^2 \|\mathbf{g}^H_v \mathbf{\Phi}\|^2$.

In Eq. \eqref{equ:6}, we adopt an approximation given by 
$\mathbb{E}\left\{\log_2\left(1+\frac{x}{y}\right)\right\} \approx \log_2\left(1+\frac{\mathbb{E}\left\{x\right\}}{\mathbb{E}\left\{y\right\}}\right),$
for the random variables $x$ and $y$.
This approximation accuracy can be estimated using the following lemma \cite{R28}.
\begin{Lemma}
Given the random variables $x$ and $y$, we have
\begin{align}\label{equ:7}
&\left| \mathbb{E}\left\{\log_2\left(1+\frac{x}{y}\right)\right\} - \log_2\left(1+\frac{\mathbb{E}\left\{x\right\}}{\mathbb{E}\left\{y\right\}}\right) \right| \nonumber \\
&~~~\leq \log_2 \left( \left( 1+ c_1 \frac{\mathbb{V} \{x+y\}}{ \mathbb{E}^2 \{x+y\}}\right)\left( 1+ c_2 \frac{\mathbb{V} \{y\}}{ \mathbb{E}^2 \{y\}}\right)  \right)
\end{align}
when $\frac{\mathbb{V} \{x+y\}}{ \mathbb{E}^3\{x+y\}} < \delta_1$ and $\frac{\mathbb{V} \{y\}}{ \mathbb{E}^3\{y\}} < \delta_2$ for positive $c_1$, $c_2$, $\delta_1$ and $\delta_2$.

\end{Lemma}

\begin{remark}
The proof is available in \cite{R28a}. $\hfill{\blacksquare}$
\end{remark}

The closed-form expression of $\hat{R}_v$ is evaluated using the following theorem.
\begin{theorem}
The approximated ergodic rate of an active RIS-aided WPC with phase quantization $\hat{R}_v$ in Eq. \eqref{equ:6} can be written as
\begin{equation}
\hat{R}_v = \left(1-\alpha\right)\log_2 \left(1+ \frac{\nu_1(t_1 + t_2 t_3 + t_4 + t_5)}{t_6} \right)
\label{equ:8}
\end{equation}
where $t_1 = \zeta_{h_p}\zeta_{f}$,  $t_2 = \zeta_{h_p}\sqrt{\pi \zeta_{f}}$, $t_3 = \sum_{m=1}^{M} \frac{\pi \sin \tau}{4\tau} \rho_m \sqrt{\zeta_{g_m}} \sqrt{\zeta_{h_m}}$, $t_4 = \sum_{m=1}^{M} \rho_m^2 \zeta_{h_p} \zeta_{g_m} \zeta_{h_m} \left(1- \frac{\pi^2 \sin^2 \tau}{16\tau^2} \right)$, $t_5 = \left(\sum_{m=1}^{M} \frac{\pi \sin \tau}{4\tau} \rho_m \sqrt{\zeta_{h_p}}  \sqrt{\zeta_{g_m}} \sqrt{\zeta_{h_m}}\right)^2$ and $t_6 = \sigma_{v}^2 M \rho_m^2 \zeta_{g_m}+\sigma_{n}^2$.
\end{theorem}

\begin{remark}

The closed-form expression of $\mathbb{E}\left\{T_1 +T_2\right\}$ in Eq. \eqref{equ:6} can be calculated as:

\begin{align}
\mathbb{E}&\left\{T_1 +T_2\right\} \nonumber \\
&=\mathbb{E}\left\{|h_p|^2\left( |f|+  \sum_{m=1}^{M}\rho_m|g_m||h_m| \cos{\left(\phi^e_m\right)}\right)^2 \right\}\nonumber \\
&+\mathbb{E}\left\{|h_p|^2\left(  \sum_{m=1}^{M}\rho_m|g_m||h_m| \sin{\left(\phi^e_m\right)}\right)^2\right\}
\label{equ:9}
\end{align}
\normalsize
which can be simplified as:
\begin{align}
\mathbb{E}&\left\{T_1 +T_2\right\}\nonumber \\
&= \mathbb{E}\left\{|h_p|^2\right\} \mathbb{E}\left\{|f|^2\right\} + 2\mathbb{E}\left\{|h_p|^2\right\} \mathbb{E}\left\{|f|\right\}\mathbb{E}\left\{s_4\right\} \nonumber \\
&+ \mathbb{E}\left\{|h_p|^2\right\}\mathbb{E}\left\{s_5\right\} + \mathbb{E}\left\{|h_p|^2\right\} \mathbb{E}\left\{s_6\right\}
\label{equ:10}
\end{align}
where $s_4 = \sum_{m=1}^{M}\rho_m|g_m||h_m| \cos\left(\phi^e_m\right)$, $s_5 = \left( \sum_{m=1}^{M}\rho_m|g_m||h_m| \cos{\left(\phi^e_m\right)}\right)^2$ and $s_6 = \left( \sum_{m=1}^{M}\rho_m|g_m||h_m| \sin{\left(\phi^e_m\right)}\right)^2$.

Considering that $|f|$, $|h_p|$, $|g_m|$ and $|h_m|$ follow Rayleigh distributions with parameters $\frac{\zeta_{f}}{2}$, $\frac{\zeta_{p}}{2}$, $\frac{\zeta_{g_m}}{2}$ and $\frac{\zeta_{h_m}}{2}$, respectively, we can deduce their first and second moments. Specifically, the expected values (first moments) are calculated as $\mathbb{E}\left\{|f|\right\}=\frac{\sqrt{\pi \zeta_{f}}}{2}$, $\mathbb{E}\left\{|h_p|\right\}=\frac{\sqrt{\pi \zeta_{p}}}{2}$, $\mathbb{E}\left\{|g_m|\right\}=\frac{\sqrt{\pi \zeta_{g_m}}}{2}$ and $\mathbb{E}\left\{|h_m|\right\}=\frac{\sqrt{\pi \zeta_{h_m}}}{2}$. Additionally, the second moments are given by $\mathbb{E}\left\{|f|^2\right\}=\zeta_{f}$, $\mathbb{E}\left\{|h_p|^2\right\}=\zeta_{h_p}$, $\mathbb{E}\left\{|g_m|^2\right\}=\zeta_{g_m}$ and $\mathbb{E}\left\{|h_m|^2\right\}=\zeta_{h_m}$. To determine the values of $\mathbb{E}\left\{s_4\right\}$, $\mathbb{E}\left\{s_5\right\}$ and $\mathbb{E}\left\{s_6\right\}$, we need to compute $\omega_{n,m}$, which is defined as the $n$-th moment of $\rho  |g_m||h_m|$. This is expressed as: $\omega_{n,m}=\mathbb{E}\left\{ \left(\rho_m  |g_m||h_m| \right)^n \right\}$. By applying the properties of expected values and Rayleigh distributions, $\omega_{n,m}$ simplifies to $\rho^n  \zeta_{g_m}^{\frac{n}{2}} \zeta_{h_m}^{\frac{n}{2}}\left(\Gamma\left(\frac{n}{2}+1\right)\right)^2$. Using this simplification, $\mathbb{E}\left\{s_4\right\}$, $\mathbb{E}\left\{s_5\right\}$ and $\mathbb{E}\left\{s_6\right\}$ can be computed as in Eqs. \eqref{equ:11}--\eqref{equ:13}. Using Eqs. \eqref{equ:9}--\eqref{equ:13}, $\mathbb{E}\left\{T_1 +T_2\right\}$ can be derived as $t_1 + t_2 t_3 + t_4 + t_5$.

\begin{figure*}[t!]
\begin{align}
&\mathbb{E}\left\{s_4\right\}=\sum_{m=1}^{M}\omega_{1,m} \mathbb{E}\left\{\cos{\left(\phi^e_m\right)}\right\}=  \sum_{m=1}^{M} \frac{\pi \sin \tau}{4\tau} \rho_m \sqrt{\zeta_{g_m}} \sqrt{\zeta_{h_m}},\quad \quad \quad \quad \quad \quad \quad \quad \quad \quad \quad \quad \quad \quad
\label{equ:11}
\end{align}

\begin{align}
\mathbb{E}\left\{s_5\right\} &=\mathbb{V}\left\{s_4\right\}+\left(\mathbb{E}\left\{s_4\right\}\right)^2 \nonumber \\
&= \sum_{m=1}^{M}\left(\omega_{2,m} \mathbb{E}\left\{\cos^2\left(\phi^e_m\right)\right\} - \left(\omega_{1,m} \mathbb{E}\left\{\cos{\left(\phi^e_m\right)}\right\} \right)^2\right)+\left(\sum_{m=1}^{M}\omega_{1,m} \mathbb{E}\left\{\cos{\left(\phi^e_m\right)}\right\}
\right)^2\nonumber \\
&=\sum_{m=1}^{M} \rho_m^2 \zeta_{g_m} \zeta_{h_m} \left(\left(\frac{\sin\left(2\tau\right)}{4\tau}+\frac{1}{2}\right)- \frac{\pi^2 \sin^2 \tau}{16\tau^2} \right) +\left(\sum_{m=1}^{M} \frac{\pi \sin \tau}{4\tau} \rho_m \sqrt{\zeta_{g_m}} \sqrt{\zeta_{h_m}}
\right)^2,
\label{equ:12}
\end{align}

\begin{align}\label{equ:13}
\mathbb{E}\left\{s_6\right\}&=\mathbb{V}\left\{\sqrt{s_6}\right\}+ \left(\mathbb{E}\left\{\sqrt{s_6}\right\}\right)^2 \quad \quad \quad \quad \quad \quad \nonumber \\
&=\sum_{m=1}^{M} \left( \omega_{2,m}\mathbb{E} \left\{ \sin^2 \left( \phi^e_m \right) \right\}-\left( \omega_{1,m}\mathbb{E} \left\{ \sin \left( \phi^e_m \right) \right\} \right)^2 \right)   +\left( \sum_{m=1}^{M} \omega_{1,m}\mathbb{E}\left\{ \sin \left( \phi^e_m \right) \right\}\right)^2  \nonumber \\
&=\sum_{m=1}^{M} \rho_m^2 \zeta_{g_m} \zeta_{h_m} \left( \frac{1}{2}-\frac{\sin\left( 2\tau \right)}{4\tau}\right).
\end{align}
\normalsize

\vspace{5px}
\hrulefill
\end{figure*}

Furthermore, the property of a norm is utilized to express $\|\mathbf{g}^H_v \mathbf{\Phi}\|^2$, which represents the sum of $M$ i.i.d. RVs \cite{R29}. The expected value of this norm, denoted by $\mathbb{E}\left\{\|\mathbf{g}^H_v \mathbf{\Phi}\|^2\right\}$, is derived as: $\sum_{m=1}^{M}\mathbb{E}\left\{ \rho_m^2 |g_m|^2\right\}$. This simplifies  $\mathbb{E}\left\{T_3+\sigma_{n}^2\right\}$ to $\sigma_{v}^2 
 M \rho_m^2 \zeta_{g_m}+\sigma_{n}^2$. Consequently, the theorem is proved. $\hfill{\blacksquare}$
\end{remark}

\normalsize \subsection{Outage Probability}

The outage probability ($P_O$) is a critical measure in wireless communication systems, indicating the likelihood that the system's capacity will fall below the required threshold, failing to decode the transmitted signal. In mathematical terms, an outage occurs when the capacity of the channel is less than the target transmission rate ($r_v$) \cite{R31}. This can be formulated as follows:
\begin{align} 
&P_O =Pr [ \left(1-\alpha \right) \log_2 \left(1+\gamma_{v} \right) < r_v ] \nonumber \\
& = Pr \left[ \left(1-\alpha\right) \right. \nonumber \\
& \!\times \!\left. \log_2\!\left(\!1\!+\!\frac{\nu_1 |h_p|^2 \left| |f|+  \sum_{m=1}^{M}\rho_m|g_m||h_m| e^{j\phi^e_m}\right|^2}{\sigma_{v}^2\|\mathbf{g}^H_v \mathbf{\Phi}\|^2+\sigma_{n}^2} \right)\!<\!r_v\right]. 
\label{equ:14}
\end{align}
Substituting $\kappa=2^{\frac{r_v}{1-\alpha}-1}$, we obtain:
\begin{align} 
P_O=\Pr\left[\frac{\nu_1 |h_p|^2 \left| |f|+   \sum_{m=1}^{M}\rho_m|g_m||h_m| e^{j\phi^e_m}\right|^2}{\sigma_{v}^2\|\mathbf{g}^H_v \mathbf{\Phi}\|^2+\sigma_{n}^2}<\kappa\right].
\label{equ:15}
\end{align}

\begin{theorem}
The outage probability of the active RIS-aided WPC with phase quantization under the target rate of $r_v$, denoted as $P_O$, is derived as shown in Eq. \eqref{equ:16}.
\end{theorem}

\begin{remark}

To address Eq. \eqref{equ:15}, initially, we derive the cumulative distribution function (CDF) and probability density function (PDF) of $\left| |f|+  \sum_{m=1}^{M}\rho_m |g_m||h_m| e^{j\phi^e_m}\right|$. Deriving exact CDF and PDF is challenging owing to their intricacy. To simplify, we employ the Gamma distribution to achieve a close approximation \cite{R32}. Given the presence of a complex number summation in $ |f|+   \sum_{m=1}^{M}\rho_m|g_m||h_m| \cos{\left(\phi^e_m\right)}+j\sum_{m=1}^{M}\rho_m|g_m||h_m| \sin{\left(\phi^e_m\right)} $, the exact distribution is difficult to determine directly, we adopt a closely fitting approximation for our subsequent analyses.
 \begin{align} 
|f|+   \sum_{m=1}^{M}\rho_m|g_m|&|h_m| \cos{\left(\phi^e_m\right)}+j\sum_{m=1}^{M}\rho_m|g_m||h_m| \sin{\left(\phi^e_m\right)} \nonumber \\
&\geq  \underbrace{|f|+   \sum_{m=1}^{M}\rho_m|g_m||h_m| \cos{\left(\phi^e_m\right)}}_{X}
\label{equ:17}
\end{align}  \normalsize

Subsequently, we employ the Gamma distribution for $X$, i.e., $X \sim \mathcal{GM}\left(s,r\right)$. The CDF of $X$ can be defined as 
$F_X\left(x;s,r\right)=\frac{\gamma\left(s,\frac{x}{r}\right)}{\Gamma\left(s\right)}$, where $\gamma\left(s,\frac{x}{r}\right)$ is a lower incomplete gamma function \cite{R33}. Thus, we can derive the PDF of $X$ as follows,
 \begin{align}
 f_X\left(x;s,r\right)=\frac{d}{dx}F_X\left(x;s,r\right)=\frac{1}{r^s\Gamma{\left(s\right)}}x^{s-1}e^{-\frac{x}{r}}
 \label{equ:18}
\end{align}  \normalsize
where $x\geq 0$, $s> 0$ (shape parameter), $r> 0$ (scale parameter), and $\Gamma{\left(s\right)}$ denotes the gamma function evaluated at $s$.

Given the intricate relationship between $\mathbb{E}\left\{|f|+   \sum_{m=1}^{M}\rho_m|g_m||h_m| \cos{\left(\phi^e_m\right)}\right\}$ and $\mathbb{V}\left\{|f|+   \sum_{m=1}^{M}\rho_m|g_m||h_m| \cos{\left(\phi^e_m\right)}\right\}$ with $s$ and $r$ as desfined by $s=\frac{\left(\mathbb{E}\left\{|f|+   \sum_{m=1}^{M}\rho_m|g_m||h_m| \cos{\left(\phi^e_m\right)}\right\}\right)^2}{\mathbb{V}\left\{|f|+   \sum_{m=1}^{M}\rho_m|g_m||h_m| \cos{\left(\phi^e_m\right)}\right\}}$ and $r=\frac{\mathbb{V}\left\{|f|+   \sum_{m=1}^{M}\rho_m|g_m||h_m| \cos{\left(\phi^e_m\right)}\right\}}{\mathbb{E}\left\{|f|+   \sum_{m=1}^{M}\rho_m|g_m||h_m| \cos{\left(\phi^e_m\right)}\right\}}$, it becomes imperative to derive the mean and variance in the following manner:

 \begin{align} 
\mathbb{E}&\left\{|f|+\   \sum_{m=1}^{M}\rho_m|g_m||h_m| \cos{\left(\phi^e_m\right)}\right\}= \underbrace{\frac{\sqrt{\pi \zeta_{f}}}{2}}_{\mathbb{E}\left\{|f|\right\}}\nonumber \\
&+\underbrace{\sum_{m=1}^{M} \frac{\pi \sin \tau}{4\tau} \rho_m \sqrt{\zeta_{g_m}} \sqrt{\zeta_{h_m}}}_{\mathbb{E}\left\{   \sum_{m=1}^{M}\rho_m|g_m||h_m| \cos{\left(\phi^e_m\right)}\right\} }
\label{equ:19}
\end{align}  \normalsize
and
 \begin{align} 
\mathbb{V}&\left\{|f|+   \sum_{m=1}^{M}\rho_m|g_m||h_m| \cos{\left(\phi^e_m\right)}\right\}=\underbrace{\zeta_{f}-\frac{\pi \zeta_{f}}{4}}_{\mathbb{V}\left\{|f|\right\}} \nonumber \\
&+\underbrace{\sum_{m=1}^{M} \rho_m^2 \zeta_{g_m} \zeta_{h_m} \left(\left(\frac{\sin\left(2\tau\right)}{4\tau}+\frac{1}{2}\right)- \frac{\pi^2 \sin^2 \tau}{16\tau^2} \right)}_{\mathbb{V}\left\{   \sum_{m=1}^{M}\rho_m|g_m||h_m| \cos{\left(\phi^e_m\right)}\right\} }
\label{equ:20}
\end{align}  \normalsize

The outage probability $P_O$ can be computed as follows:

\begin{align} 
P_O&=\Pr\left[|h_p|^2  X^2<\frac{\kappa\left(\sigma_{v}^2 M \rho_m^2 \zeta_{g_m}+\sigma_{n}^2\right)}{\nu_1}\right] \nonumber \\
\label{equ:21aa}
\end{align}\normalsize

This can be further transformed into an integral form as:

\begin{align} 
P_O&=\int_{0}^{\infty} F_{|h_p|^2}\left(\frac{\kappa\left(\sigma_{v}^2 M \rho_m^2 \zeta_{g_m}+\sigma_{n}^2\right)}{\nu_1 t^2}\right) f_X\left(t\right)dt \nonumber \\
&=1-\int_{0}^{\infty} e^{-\frac{\kappa\left(\sigma_{v}^2 M \rho_m^2 \zeta_{g_m}+\sigma_{n}^2\right)}{\nu_1 t^2}} \frac{t^{s-1}}{r^s\Gamma{\left(s\right)} }e^{-\frac{t}{r}}dt
\label{equ:21}
\end{align}\normalsize

Deriving an exact closed-form expression for $P_O$ in Eq. \eqref{equ:21} is challenging. However, this can be addressed by using the Gaussian--Chebyshev quadrature method \cite{R34} and the variable transformation $t=\tan\left( \frac{\pi}{4} \left( \cos \left( \frac{2u-1}{2U}\pi \right)+1\right) \right)$ \cite{R35}, where $U$ represents a parameter that balances accuracy against computational complexity. This method enables a precise approximation of $P_O$, which is presented as Eq. \eqref{equ:16}. Therefore, the theorem is proved.
$\hfill{\blacksquare}$
\end{remark}

\begin{figure*}[t]
\begin{align} \label{equ:16}
&P_O= 1- \nonumber \\
&\left\{ \frac{\pi^2}{U} \sum_{u=1}^{U}\sqrt{ \!1 \!- \!\left(\cos\left(\frac{2u-1}{2U}\pi \right) \right)^2}   \! \sec^2   \! \left(0.25\pi  \! \left(\cos\left(\frac{2u-1}{2U}\pi\right)+1\right)  \right)  \! \left( \! \tan  \! \left(\frac{\pi\left(\cos\left(\frac{2u-1}{2U}\pi\right)+1\right)}{4} \right)\right)^{s-1}  \! \left(r^s\Gamma{\left(s\right)}\right) ^{-1}  \right.\nonumber \\
&  \left. \exp \left[ -\kappa \left( \sigma_{v}^2 M \rho_m^2 \zeta_{g_m}+\sigma_{n}^2 \right) \left( \nu_1 \tan^2 \left( \frac{\pi \left(\cos \left(\frac{2u-1}{2U}\pi \right)+1 \right)}{4} \right) \right)^{-1} 
- \tan  \left( \frac{\pi \left(\cos \left(\frac{2u-1}{2U}\pi \right)+1 \right)}{4} \right)r^{-1} \right] \right\}
\end{align}

\vspace{10px}
\hrulefill
\end{figure*}

\subsection{Power Consumption of an Active RIS}

The power consumption of an active RIS, accounting for the amplification and transmission of the incident signal, can be expressed as follows:

\begin{align}
P_c\left(\alpha\right)=\nu_1 \|\mathbf{\Theta}\mathbf{h}_v\|^2+\sigma_{v}^2\|\mathbf{\Theta}\|^2+M\left(P_{1}+P_{2}\right),
\label{equ:22}
\end{align}
where $P_{1}$ and $P_{2}$ are constants representing the power consumed by the switching/controlling circuitry and the power required for DC biasing, respectively \cite{R36}. 

As demonstrated in Eq. \eqref{equ:22}, the power consumption in an active RIS is determined by the total power of the reflected signal and depends on the parameter $\alpha$, considering $\nu_1$ to be a function of $\alpha$. This equation provides a comprehensive model for evaluating the power efficiency of active RIS in WPC systems, highlighting the critical role of amplification and noise management in optimizing system performance.

\section{Optimizing Ergodic and Effective Rates} \label{Sec4}

Building on the analytical results, including the ergodic rate and outage probability from the previous section that incorporated phase quantization errors and active RIS amplification, this section explores the optimization of the time-switching factor $\alpha$. It examines scenarios both with and without the power consumption of an active RIS, underscoring its vital role in shaping the performance of active RIS-aided WPC systems.

\subsection{Optimizing Ergodic Rate ($\hat{R_{v}}$) with $\alpha$}

First, we consider the optimization of the ergodic rate, which is influenced by the time-switching factor $\alpha$. This factor determines the allocation between the energy harvesting and information transmission periods. Specifically, dedicating more time to energy harvesting enhances the available transmission power but shortens the time for information transmission. This introduces a tradeoff in achieving an optimal balance between energy harvesting and information transmission. Thus, we mathematically formulate the optimization problem for the ergodic rate $\hat{R_{v}}\left(\alpha\right)$ in terms of $\alpha$ as follows:

\begin{subequations} \label{equ:23}
\begin{align}
 & \max \quad \hat{R_{v}}\left(\alpha\right)=\left(1-\alpha\right)\mathbb{E}\{\log_2\left(1+\gamma_{v}\right)\}  \\ 
&    s.t. \quad \quad 0 \leq \alpha \leq 1,
 \end{align}
 \end{subequations} 
 where $\hat{R_{v}}$ represents the ergodic rate as a function of the time-switching factor $\alpha$.  The constraint (24b) ensures that $\alpha$ lies between 0 and 1, where  $\alpha=0$ corresponds to no time allocated to energy harvesting and $\alpha=1$ corresponds to no time allocated to information transmission.

To determine the optimal time-switching factor $\alpha^*$, we calculate the derivative of $\hat{R_{v}}\left(\alpha\right)$, as given by Eq. \eqref{equ:24}. The cases $\alpha = 0 $ and $\alpha= 1$ are not considered viable solutions because they lead to $\hat{R_{v}}\left(\alpha\right)= 0$, yielding trivial results. In accordance with the Karush--Kuhn--Tucker (KKT) conditions, the solution to Eq. \eqref{equ:24} is the $\alpha^*$ that satisfies $\dv{\alpha}\hat{R_{v}}\left( \alpha^* \right) =0$. Eq. \eqref{equ:24} is characterized by logarithmic and polynomial components. As a result, it introduces complexity in the analysis of the closed-form solution of $\alpha^*$ that satisfies $\dv{\alpha}\hat{R_{v}}\left(\alpha^*\right)=0$. Therefore, we employ the numerical methods in Section V to determine $\alpha^*$ such that $\dv{\alpha}\hat{R_{v}}\left(\alpha^*\right)=0$.

 \begin{figure*}[h!]
\begin{align}
\dv{\alpha}&\hat{R_{v}}\left(\alpha\right) =\frac{1}{\ln\left(2\right)}\left(\dfrac{\left(1-\alpha\right)\left(\frac{t_7\alpha}{t_6\cdot\left(1-\alpha\right)^2}+\frac{t_7}{t_6\cdot\left(1-\alpha \right)}\right)}{\left(\frac{t_7\alpha}{t_6\cdot\left(1-\alpha\right)}+1\right)}-\ln\left(\frac{t_7\alpha}{t_6\cdot\left(1-\alpha\right)}+1\right)\right)
\label{equ:24}
 \end{align}
where $t_7= \eta P_p \left( t_1+t_2t_3+t_4+t_5 \right) $. 

\vspace{5px}
\hrulefill
 \end{figure*}

\subsection{Optimizing Ergodic Rate ($\hat{R_{v}}$) Considering RIS Power Consumption}

In the previous problem, the power consumption of an active RIS was not considered based on the assumption that the RIS is connected to an inexhaustible power source, such as the power supply of a building. This assumption is valid for stationary deployment scenarios. However, in mobile RIS deployments, which are subject to power limitations from portable power sources or energy harvesting, power consumption becomes a critical constraint. Consequently, this subsection introduces and formulates an optimization problem under the constraint of the active RIS power consumption, detailed as follows:

\begin{subequations} \label{equ:25}
\begin{align}
 &\max  \quad  \hat{R_{v}} \left(\alpha\right)=\left(1-\alpha\right)\mathbb{E}\{\log_2\left(1+\gamma_{v}\right)\}  \\
& s.t.  \quad  \;  \; P_c  \left(\alpha\right) \leq P_{R}, \\
& \quad \quad \quad 0 \leq \alpha \leq 1,
\end{align}
\end{subequations}
where the constraint $P_c  \left(\alpha\right) \leq P_{R}$ ensures that the power consumption of the active RIS does not exceed the maximum power consumption threshold $P_{R}$.

\begin{theorem}
The optimal time-switching factor that maximizes the ergodic rate under an RIS power consumption constraint, that is, the solution to the optimization problem \eqref{equ:25}, is given as:
The solution is $\alpha^*$ such that $\dv{\alpha}\hat{R_{v}} \left(\alpha^*\right) = 0$ if $P_c(\alpha^*) < P_R$. Otherwise, the solution is $\alpha^*$ such that $ P_c(\alpha^*) =  P_R$.
\end{theorem}

\begin{remark}
To systematically address this constrained optimization problem, we employ the KKT conditions. To this end, the Lagrangian function is defined as,

\begin{equation}
    \mathcal{L} = \hat{R_{v}} \left(\alpha\right) + \lambda_1 \left( P_c(\alpha)-P_R \right) - \lambda_2 \alpha + \lambda_3 ( \alpha -1),
\end{equation}
where $\lambda_1$, $\lambda_2$ and $\lambda_3$ are Lagrange multipliers.
The solution $\alpha^*$ to the optimization problem satisfies the following conditions:
\begin{itemize}
    \item \textit{Stationarity}: $\dv{\alpha}\mathcal{L} \left( \alpha^* \right) = \dv{\alpha}\hat{R_{v}} \left(\alpha^*\right) + \lambda_1   \dv{\alpha}P_c (\alpha^*) - \lambda_2 + \lambda_3 = 0$ 
    \item \textit{Primal Feasibility}: $P_c(\alpha^*) - P_R \leq 0$, $-\alpha^* \leq 0$ and $ \alpha^* - 1 \leq 0$
    \item \textit{Dual Feasibility}: $\lambda_1, \lambda_2, \lambda_3 \geq 0$
    \item \textit{Complementary Slackness}: $\lambda_1 (P_c(\alpha^*) - P_R) = 0$, $\lambda_2 \alpha^* = 0$ and $\lambda_3 (\alpha^*-1) = 0$
\end{itemize}

In the complementary slackness condition, $\alpha^* = 0$ and $1$ are trivial solutions because there are no energy harvesting and information transmission phases, respectively. Thus, $\lambda_2 = \lambda_3 = 0$. To address the remaining complementary slackness condition, we consider two cases: $P_c(\alpha^*) < P_R$ and $P_c(\alpha^*) = P_R$. If $P_c(\alpha^*) < P_R$, $\lambda_1 = 0$, the solution is $\alpha^*$ such that $\dv{\alpha} \hat{R_{v}} \left(\alpha^*\right) = 0$. Otherwise, $P_c(\alpha^*) = P_R$, $\lambda_1 \geq 0$. To meet the complementary slackness condition, $\lambda_1 = - \frac{ \dv{\alpha} \hat{R_{v}} \left(\alpha^*\right)}{  \dv{\alpha} P_c (\alpha^*) }$, and the solution is given by $\alpha^* = P_c^{-1} (P_R)$, where $P_c^{-1}(\cdot)$ denotes the inverse function of $P_c(\alpha)$ with respect to $\alpha$. $\hfill{\blacksquare}$
\end{remark}

\subsection{Optimizing Effective Rate ($R_{eff}$) with $\alpha$}

The effective rate $R_{eff}$ is a measure of the transmission rate that can be achieved without experiencing an outage. Given a target transmission rate $r_v$ and the probability of an outage $P_O$, the effective rate can be defined as:

\begin{equation}
    R_{eff} = \left[ 1-P_O \right] r_v.
\end{equation}
where $P_O$ is the probability that the transmission fails (outage occurs). The factor $[1-P_O]$ represents the probability of successful transmission.

To optimize the effective rate $R_{eff}$ with respect to the parameter $\alpha$, we need to solve the following optimization problem:

\begin{subequations} \label{equ:29}
\begin{align}
 &\max \quad  R_{eff}\left(\alpha\right)=\left[1-P_O \left(\alpha\right) \right]r_v \\
&   s.t. \quad \quad  0 \leq \alpha \leq 1,
 \end{align}
 \end{subequations}
 \normalsize

\begin{theorem}
 The optimal value of $\alpha$ ($\alpha^\dag$) for maximizing the effective rate based on the outage probability can be characterized as:
 
 \begin{align}
\alpha^\dag=\dfrac{1}{\ln\left(2\right)\,r_v +1}.
\label{equ:30}
\end{align}
 \end{theorem}
 
\begin{remark}

To derive the optimal value of $\alpha$, we start by computing the derivative of $R_{eff}\left(\alpha\right)$ with respect to $\alpha$. This derivative is given by $\dv{\alpha} R_{eff}(\alpha) = r_v \left[ - \dv{\alpha} P_O (\alpha) \right] $ and is computed in Eq. \eqref{equ:31}. According to the KKT conditions, the solution to Eq. \eqref{equ:31} is the $\alpha^\dag$ that satisfies $ \left[1-\dv{\alpha}P_O \left(\alpha^\dag\right) \right]r_v=0$, thereby validating the theorem.
$\hfill{\blacksquare}$

\begin{figure*}[h!]
\begin{align}
\dv{\alpha} R_{eff}&=\dv{\alpha}\left(S_7\exp\left(-\frac{S_8\left(1-\alpha\right)}{\alpha}2^{\left(\frac{r_v}{1-\alpha}-1\right)}-S_9\right)\right)\nonumber \\
&=S_7\exp\left(-\frac{S_8\left(1-\alpha\right)}{\alpha}2^{\left(\frac{r_v}{1-\alpha}-1\right)}-S_9\right)\dv{\alpha}\left(-\frac{S_8\left(1-\alpha\right)}{\alpha}2^{\left(\frac{r_v}{1-\alpha}-1\right)}-S_9\right)\nonumber \\
&=S_7\left(-\dfrac{\ln\left(2\right)\,r_v S_8 2^{\frac{r_v}{1-\alpha}-1}}{\left(1-\alpha\right)\alpha}+\dfrac{S_8 2^{\frac{r_v}{1-\alpha}-1}}{\alpha}+\dfrac{S_8\left(1-\alpha\right)2^{\frac{r_v}{1-\alpha}-1}}{\alpha^2}\right)\exp\left(-\frac{S_8\left(1-\alpha\right)}{\alpha}2^{\left(\frac{r_v}{1-\alpha}-1\right)}-S_9\right)\nonumber \\
&=\dfrac{S_7S_8 \left(\left(\ln\left(2\right)\,r_v +1\right)\alpha-1\right) 2^{\frac{r_v}{1-\alpha}-1}\exp\left(-\frac{S_8\left(1-\alpha\right)}{\alpha}2^{\left(\frac{r_v}{1-\alpha}-1\right)}-S_9\right)}{\left(\alpha-1\right)\alpha^2}
\label{equ:31}
\end{align}\normalsize\ where $S_7=r_v \frac{\pi^2}{U} \sum_{u=1}^{U}\sqrt{1-\left(\cos \left(\Upomega\right)\right)^2}\sec^2 \left(0.25\pi\left(\cos\left(\Upomega\right)+1\right) \right)\left(\tan\left(\frac{\pi\left(\cos\left(\Upomega\right)+1\right)}{4} \right)\right)^{s-1} \left(r^s\Gamma{\left(s\right)} \right)^{-1}$, 

$\Upomega= \frac{2u-1}{2U}\pi$, 

$S_8=\left(\sigma_{v}^2 M \rho_m^2 \zeta_{g_m}+\sigma_{n}^2\right)\left(\eta P_p \tan^2\left(\frac{\pi\left(\cos\left(\Upomega\right)+1\right)}{4} \right) \right)^{-1}$, 

$S_9=\tan\left(\frac{\pi\left(\cos\left(\Upomega\right)+1\right)}{4} \right)r^{-1}$. \\

\vspace{5px}
\hrulefill
 \end{figure*}
\end{remark}

\subsection{Optimizing $R_{eff}$ Considering RIS Power Consumption}

The maximization problem for $R_{eff}$ under the constraint of active RIS power consumption can be defined as:

\begin{subequations} \label{equ:30}
\begin{align}
 &\max \quad  R_{eff} \left(\alpha\right) =\left[1-P_O \left(\alpha\right) \right]r_v, \\
& s.t.  \quad \; \; P_c  \left(\alpha\right) \leq P_{R}, \\
& \quad \quad \quad 0 \leq \alpha \leq 1,
\end{align}
\end{subequations}

Following the established approach in Section~\ref{Sec4}-B, we address the optimization problem \eqref{equ:30}. Utilizing the KKT framework, we identify the optimal $\alpha$ $\left(\alpha^\dag\right)$ that maximizes $R_{eff}$. The approach involves stationarity, primal and dual feasibility, and complementary slackness conditions, with trivial solutions at $\alpha^\dag = 0$ and $1$ eliminated by setting $\lambda_2 = \lambda_3 = 0$. The optimal solution $\alpha^\dag$ depends on whether $P_c(\alpha^\dag)$ is less than or equal to $P_R$. If $P_c(\alpha^\dag) < P_R$, implying $\lambda_1 = 0$, and the optimal $\alpha^\dag$ is determined by setting $\dv{\alpha}{R_{eff}} \left(\alpha^\dag\right) = 0$. Conversely, if $P_c(\alpha^\dag) = P_R$, then $\lambda_1 \geq 0$, and $\alpha^\dag$ aligns with the power consumption constraint being actively met.

\section{Numerical Results} \label{Sec5}

In this Section, we present numerical results to validate the theoretical analysis discussed in Sections~\ref{Sec3} and \ref{Sec4}. Unless otherwise specified, we adopt the following settings: maximum amplitude gain $\rho_m^{*}$ = 6, rate constraint $r_v$ = 2 bits/s/Hz, transmission power of the wireless RF power hub $P_p$ = 20 dB, power consumption threshold of the active RIS $P_{R}$ = 10 mW, noise power at the active RIS $\sigma_{v}^2$ = -80 dBm, noise power at the $V_2$ $\sigma_{n}^2$ = -80 dBm, $d_p$ = 20 m,  $d_f$ = 30 m,  $d_{h_m}$ = $d_{g_m}$ = 20 m, path loss exponent $\epsilon$ = 3, number of active elements $M$ = 36, time switching factor $\alpha$ = \{0.1, 0.9\}, $P_1$ = $P_2$ = -10 dBm, $\eta$ = 0.8, $\tau_c$ = 1 and $b$ = $\{1, 4\}$. Monte Carlo simulations are performed over $10^6$ channel realizations to obtain the simulated results \cite{R37}. The numerical results are represented by curve lines, whereas the dot markers indicate the simulation results. Notably, the negligible discrepancy between the simulation and numerical results substantiates the validity of our theoretical analysis.

\begin{figure}[h!]
\centering
\includegraphics[width=1\linewidth]{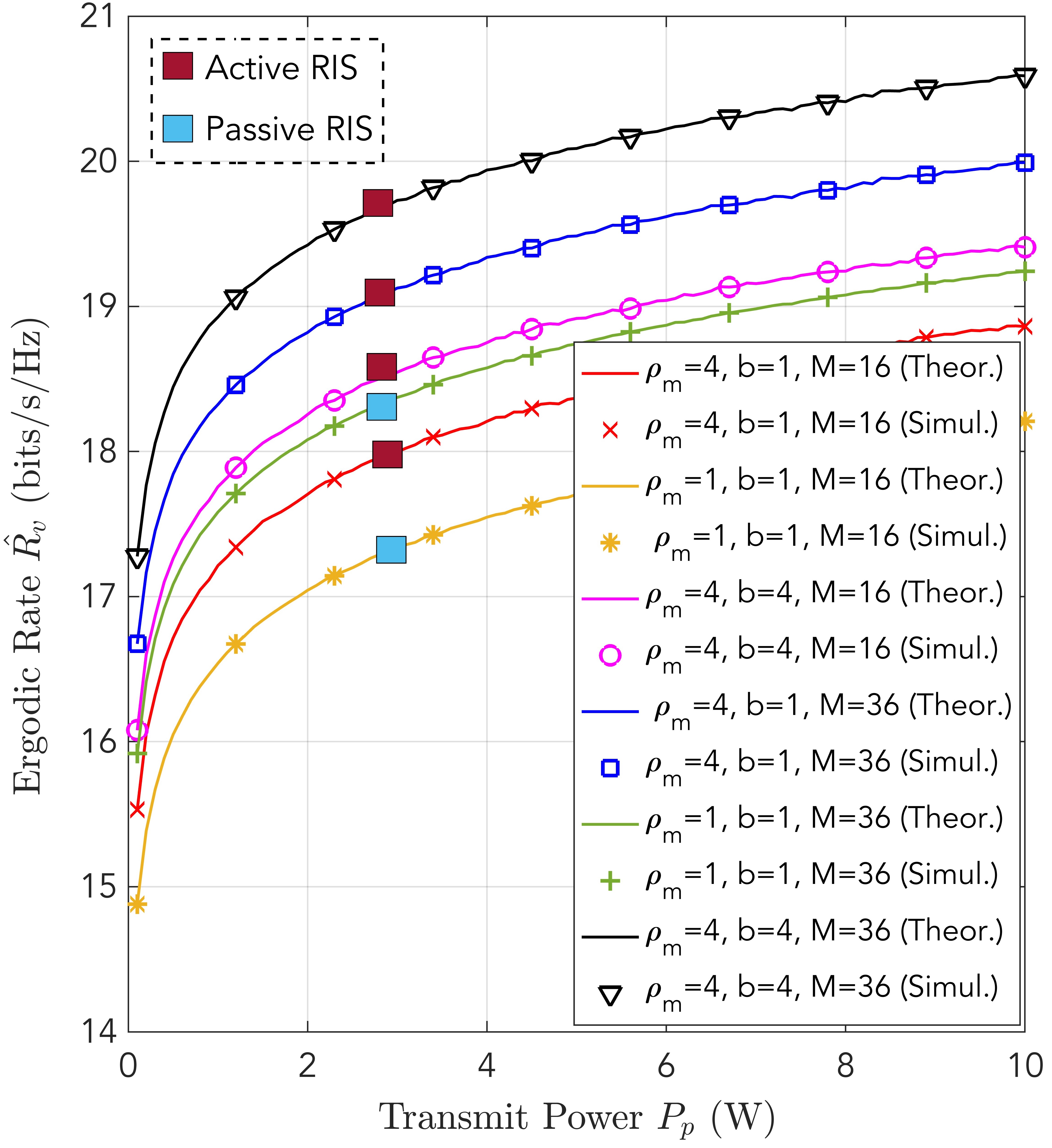}
\caption{Ergodic Rate ($\hat{R_{v}}$) vs. transmission power of a wireless RF power hub $P$ ($P_p$).}
\label{fig2}
\end{figure}

Fig.~\ref{fig2} illustrates the ergodic rate $\hat{R_{v}}$ as a function of the transmission power of the wireless RF power hub ($P$), $P_p$. The results highlight the superior performance of an active  RIS compared to a passive RIS, even in the presence of phase quantization errors. The ergodic rate for an active RIS without phase quantization errors, {serving} as a benchmark for optimal performance, is also depicted. In our analysis, $b=4$ represents an ideal phase shifting scenario, consistent with the findings in \cite{R38} that a 4-bit quantized RIS system closely approximates the performance of continuous phase shifting. The analysis reveals that, although an active RIS introduces non-negligible RIS-response-dependent noise, its signal amplification capability substantially mitigates the adverse effects of double fading, thereby enhancing the ergodic rate. Consequently, an active RIS can achieve performance gains comparable to those of a passive RIS but with significantly fewer reflecting elements.

\begin{figure}[h!]
\centering
\includegraphics[width=1\linewidth]{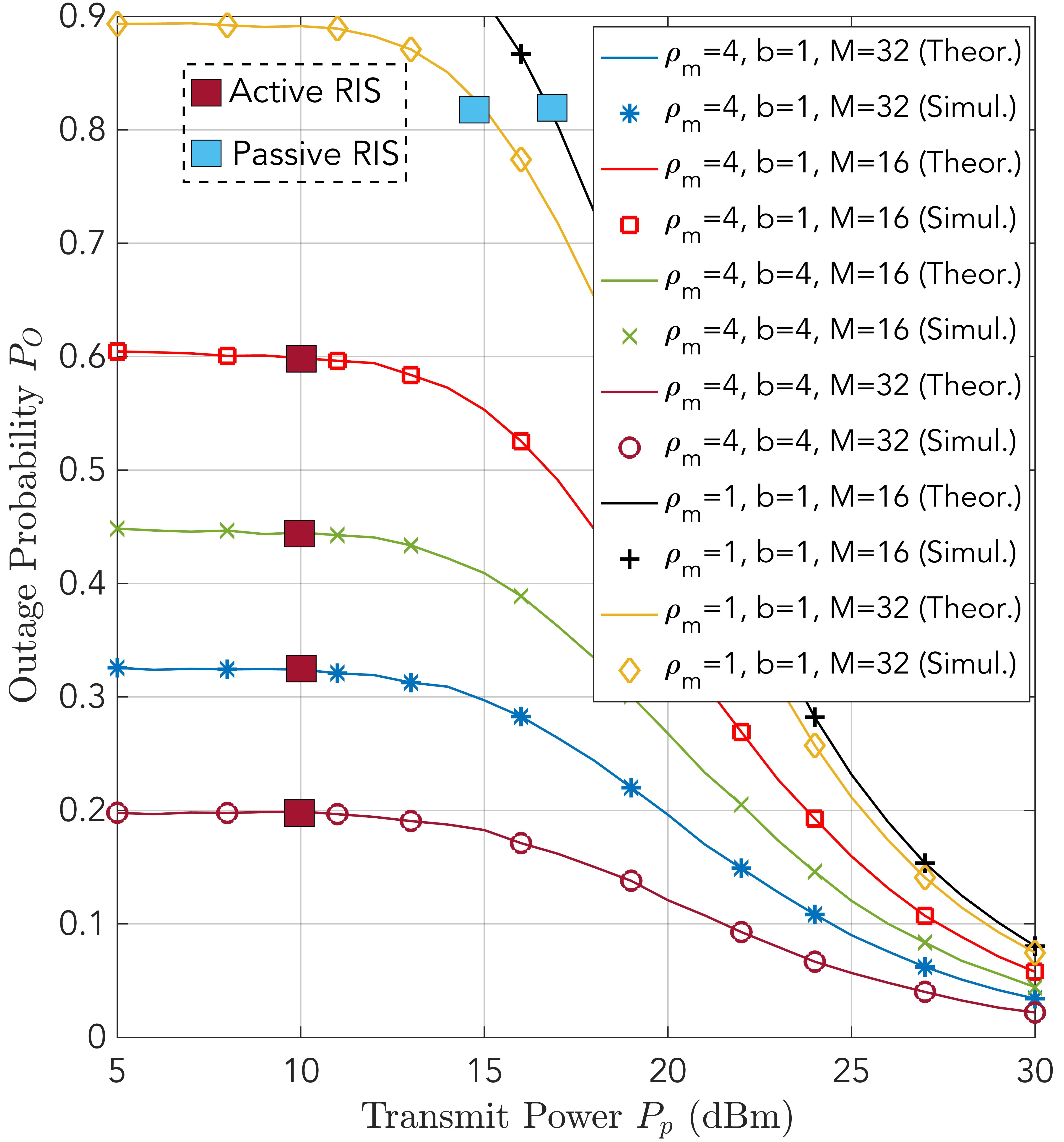}
\caption{Outage Probability ($P_O$) vs. transmission power of a wireless RF power hub $P$ ($P_p$).}
\label{fig3}
\end{figure}

Fig.~\ref{fig3} presents the outage probability $P_O$ as a function of the transmission power $P_p$ for varying numbers of reflecting elements. This analysis underscores the benefits of increased reflecting elements and the detrimental impact of phase quantization errors. Specifically, the results compare the outage performance of active and passive RISs under phase quantization errors, including a performance evaluation of an active RIS without these errors. The findings confirm the superior outage performance of an active RIS. For instance,  at $P_p$ = 20 dBm, an active RIS achieves the same outage probability as a passive RIS with half the number of reflecting elements, requiring only 16 instead of 32.

\begin{figure}[h!]
\centering
\includegraphics[width=1\linewidth]{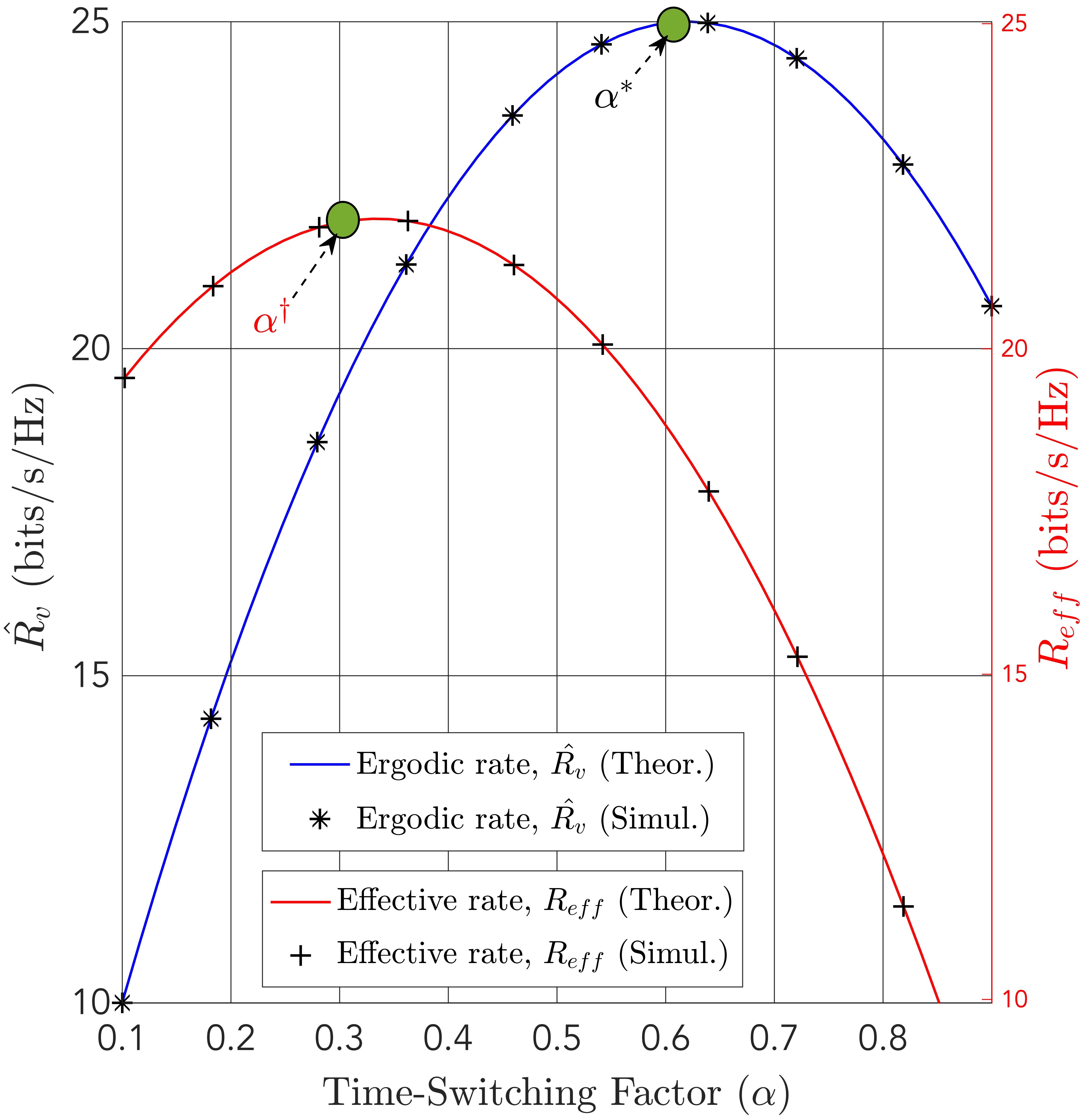}
\caption{Ergodic rate ($\hat{R_{v}}$) and effective rate ($R_{eff}$) vs. time switching factor $\alpha$.}
\label{fig4}
\end{figure}

Fig.~\ref{fig4} depicts the relationship between the time-switching factor $\alpha$ and both the ergodic rate $\hat{R_{v}}$ and the effective rate $R_{eff}$. The values for $\hat{R_{v}} (\alpha)$ and $R_{eff} (\alpha)$ are computed using Eqs. \eqref{equ:23} and \eqref{equ:29}, respectively. The results determine the optimal points for $\hat{R_{v}} (\alpha)$ and $R_{eff} (\alpha)$, denoted as $\alpha^*$ and $\alpha^\dag$, respectively. These optimal points are crucial for optimizing an active RIS-enhanced WPC system, demonstrating a balance between energy consumption and efficient data transmission utilizing the harvested energy.

\begin{figure}[h!]
\centering
\includegraphics[width=1\linewidth]{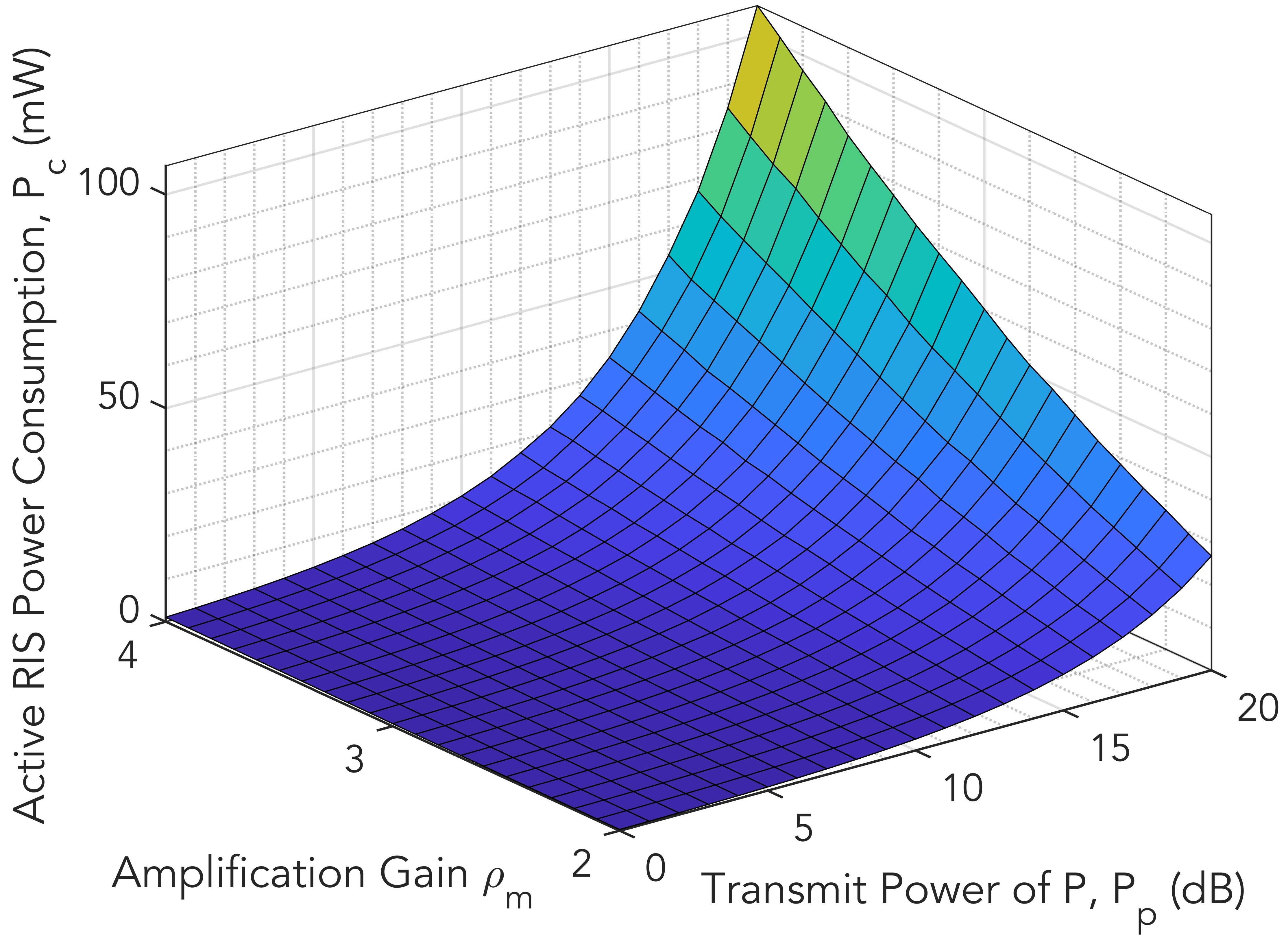}
\caption{Power consumption of an active RIS ($P_{c}$) vs. amplification gain ($\rho_{m}$) and transmission power of $P$ ($P_{p}$).}
\label{fig5}
\end{figure}

\begin{figure}[h!]
\centering
\includegraphics[width=1\linewidth]{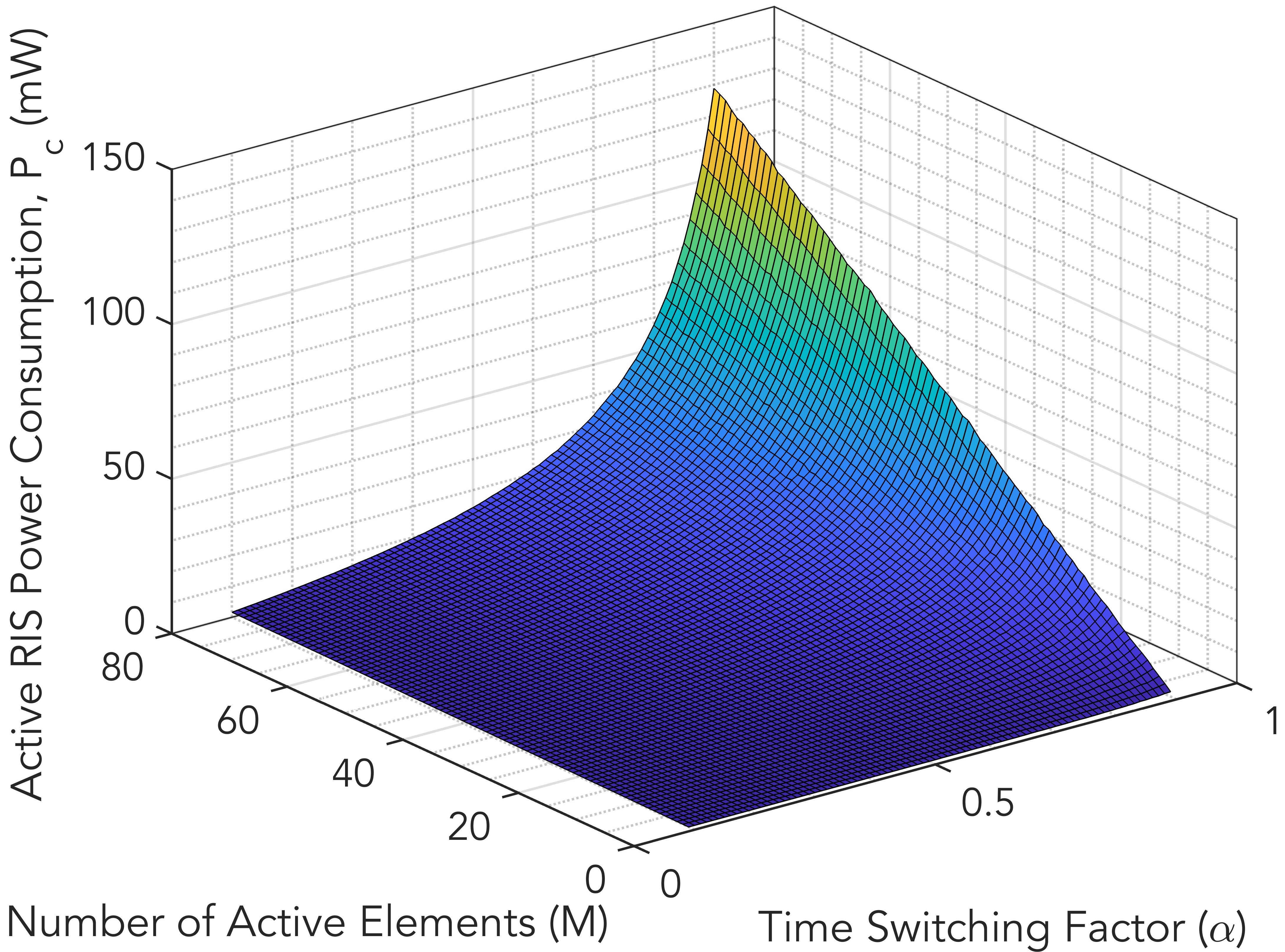}
\caption{Power consumption of an active RIS ($P_{c}$) vs. number of active RIS elements ($M$) and time switching factor ($\alpha$).}
\label{fig6}
\end{figure}

Figs.~\ref{fig5} and \ref{fig6} illustrate the power consumption of an active RIS $P_{c}$. The data points are derived from Eq. \eqref{equ:22}. Importantly, $P_{c}$ is examined in relation to the amplification gain $\rho_{m}$, transmit power $P_{p}$, number of active elements $M$, and time-switching factor $\alpha$. Notably, $M$ emerges as the most significant factor affecting $P_c$, followed by $\rho_{m}$ and $\alpha$. This detailed analysis highlights the interplay between the physical attributes of an active RIS and its power efficiency. Such insights are invaluable for the strategic design and optimization of WPC systems to ensure enhanced energy transfer and data transmission capabilities.

\section{Conclusion} \label{Sec6}
In this article, we investigated WPC systems for IoT devices with limited energy, focusing on the application of RIS technology to mitigate blockages and enhance wireless connectivity. We focused on active RIS to address double-fading attenuation by leveraging its power amplification capabilities for efficient signal transmission. We comprehensively examined the adjustable parameters of an active RIS within WPC, including signal amplification, active noise, power consumption, and phase quantization errors, to assess the ergodic rate and outage probability. Additionally, we optimized WPC scenarios by considering the time-switching factor and active RIS power consumption. The results validated our analytical framework and demonstrated significant performance improvements of active RIS over passive RIS.

Future work will involve integrating machine learning with active RIS-enabled WPC systems to optimize energy efficiency and adapt to dynamic environments. We also plan to explore multi-RIS configurations to enhance network coverage and reliability in complex IoT ecosystems. Moreover, we will address practical challenges for active RIS, including imperfect CSI, self-interference, and processing delays.

\begin{IEEEbiography}[{\includegraphics[width=1in,height=1.25in,clip,keepaspectratio]{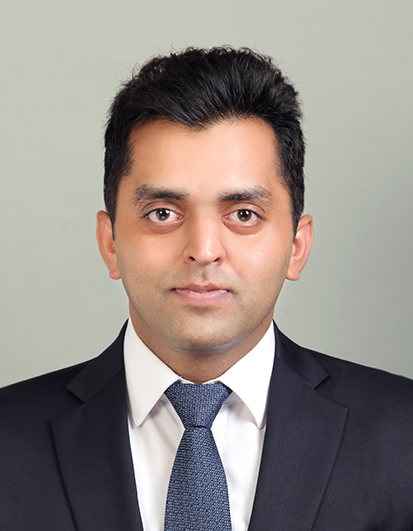}}]{Waqas Khalid} received the B.S. degree in Electronics Engineering from GIK Institute of Engineering Sciences and Technology, KPK, Pakistan, in 2011. He received M.S. and Ph.D. degrees in information and communication engineering from Inha University, Incheon, South Korea, and Yeungnam University, Gyeongsan, South Korea, in 2016, and 2019, respectively. He is currently working as a research professor at the Institute of Industrial Technology, Korea University, Sejong, South Korea. His areas of interest include physical layer modeling, signal processing, and emerging technologies for 5G networks, including reconfigurable intelligent surfaces, physical-layer security, NOMA, cognitive radio, UAVs, and IoTs. \end{IEEEbiography}

\begin{IEEEbiography}[{\includegraphics[width=1in,height=1.25in,clip,keepaspectratio]{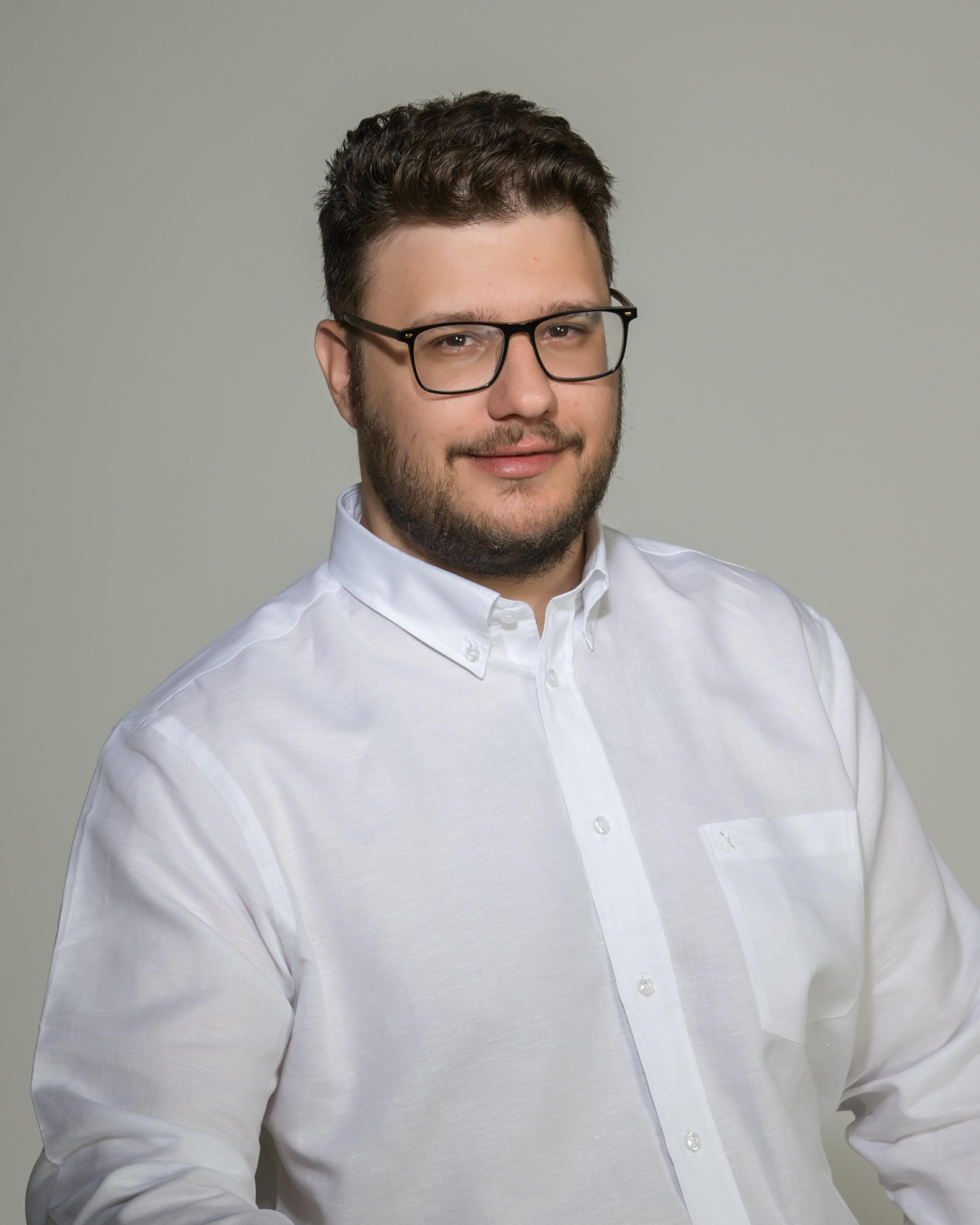}}] {Alexandros-Apostolos A. Boulogeorgos } (Senior Member, IEEE) received the Diploma degree in Electrical and Computer Engineering and the Ph.D. degree in wireless communications from the Aristotle University of Thessaloniki in 2012 and 2016, respectively. From 2022, he is an Assistant Professor at the Department Electrical and Computer Engineering of the University of Western Macedonia, Greece. Dr Boulogeorgos has (co-)authored more than 160 technical papers, which were published in scientific journals and presented at prestigious international conferences. Furthermore, he is the holder of 2 (1 national and 1 European) patents, while he has filled other 3 patents. He is listed in ``World’s Top 2\% Scientists for the Year 2022'' which is published by Stanford University and Elsevier. His current research interests spans in the area of wireless communications and networks with emphasis in high frequency communications, intelligent communication systems with emphasis to semantic communications, optical wireless communications, and signal processing and communications for biomedical applications. He has been involved as a member of organizational and technical program committees in several IEEE and non-IEEE conferences and served as a reviewer in various IEEE journals and conferences. He is an IEEE Senior Member and a Member of the Technical Chamber of Greece. He is currently an Editor for IEEE TRANSACTIONS ON WIRELESS COMMUNICATIONS, IEEE COMMUNICATIONS LETTERS, Frontier in Communications and Networks, and MDPI~Telecom. Dr Boulogeorgos has participated as a guest editor in the organization of a number of special issues in IEEE and non-IEEE journals. \end{IEEEbiography}

\begin{IEEEbiography}[{\includegraphics[width=1in,height=1.25in,clip,keepaspectratio]{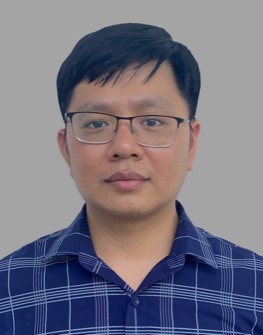}}]{Trinh Van Chie} (Member, IEEE)  received the B.S. degree in electronics and telecommunications from the Hanoi University of Science and Technology (HUST), Vietnam, in 2012, the M.S. degree in electrical and computer engineering from Sungkyunkwan University (SKKU), South Korea, in 2014, and the Ph.D. degree in communication systems from Linköping University (LiU), Sweden, in 2020. He was a Research Associate with the University of Luxembourg. He is currently with the School of Information and Communication Technology (SoICT), HUST, Vietnam. His research interests include convex optimization problems and machine learning applications for wireless communications, and image and video processing. He also received the Award of Scientific Excellence in the first year of the 5G Wireless Project funded by European Union Horizon’s 2020. He was an Exemplary Reviewer of IEEE WIRELESS COMMUNICATIONS LETTERS in 2016, 2017, and 2021, and IEEE TRANSACTONS ON COMMUNICATIONS in 2022. \end{IEEEbiography} 

\begin{IEEEbiography}[{\includegraphics[width=1in,height=1.25in,clip,keepaspectratio]{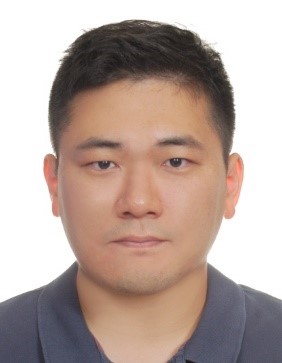}}]{Junse Lee} received the B.S. and M.S. degrees in electrical engineering from KAIST, Daejeon, South Korea, in 2009 and 2011, respectively, and the Ph.D. degree in electrical and computer engineering from the University of Texas at Austin, Austin, TX, USA, in 2018. He is currently an Assistant Professor with the School of AI Convergence, Sungshin Women's University, Seoul, South Korea. Before joining Sungshin Women's University, he was a staff engineer at Samsung Electronics. His research interests include the modeling and analysis of wireless networks. \end{IEEEbiography} 

\begin{IEEEbiography}[{\includegraphics[width=1in,height=1.25in,clip,keepaspectratio]{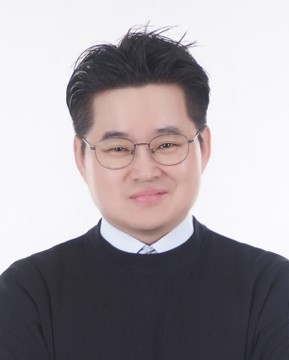}}]{Howon Lee} (S'04-M'12-SM'22) received the B.S., M.S., and Ph.D. degrees in Electrical and Computer Engineering from the Korea Advanced Institute of Science and Technology (KAIST), Daejeon, South Korea, in 2003, 2005, and 2009, respectively. From 2009 to 2012, he was a senior research staff/team leader of the knowledge convergence team at the KAIST Institute for Information Technology Convergence (KI-ITC). From 2012 to 2024, he was with the School of Electronic and Electrical Engineering and the Institute for IT Convergence (IITC) at Hankyong National University (HKNU), Anseong, South Korea. Since 2024, he has been with the Department of Electrical and Computer Engineering at Ajou University, Suwon, South Korea. He has also experienced as a Visiting Scholar with the University of California, San Diego (UCSD), La Jolla, CA, USA, in 2018. His current research interests include B5G/6G wireless communications, ultra-dense distributed networks, in-network computations for 3D images, cross-layer radio resource management, reinforcement learning for UAV and satellite networks, unsupervised learning for wireless communication networks, and Internet of Things.\end{IEEEbiography} 

\begin{IEEEbiography}[{\includegraphics[width=1in,height=1.25in,clip,keepaspectratio]{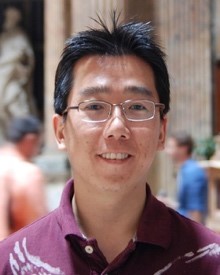}}]{Heejung Yu} (S'07-M'12-SM'20) received the B.S. degree in Radio Science and Engineering from Korea University, Seoul, Korea, in 1999 and the M.S. and Ph. D. degrees in Electrical Engineering from the Korea Advanced Institute of Science and Technology (KAIST), Daejeon, Korea, in 2001 and 2011, respectively. From 2001 to 2012, he has been with the Electronics and Telecommunications Research Institute (ETRI), Daejeon, Korea. From 2012 to 2019, he was with Yeungnam University, Korea. Currently, he is a Professor in the Department of Electronics and Information Engineering, Korea University, Sejong, Korea. His areas of interest include statistical signal processing and communication theory. \end{IEEEbiography} 

\end{document}